\documentstyle[12pt]{article}
\begin{document}

{\Large 
\centerline {Wave channeling of X-rays in narrow
rough capillaries -} 
\centerline {non Andronov-Leontovich theory.}
}

\vskip 15pt
{
\centerline {L.I.Ognev\footnote[1]{
Nuclear Fusion Institute, 
Russian Research Center "Kurchatov
Institute", Moscow, 123182, Russia \\
E-mail: ognev@qq.nfi.kiae.su
}
}


\par

{\abstract
The effect of capture of X-ray beam into narrow submicron capillary
was investigated with account for diffraction and decay of coherency
by roughness scattering in transitional boundary layer. 
In contrast to well-known Andronov-Leontovich approach the losses
do not vanish at zero gliding angle and scale proporpional to the
first power of roughness amplitude for small gliding angles.
It was shown that for small correlation radius of roughness the
scattering decay of coherency can be made of the same order as
absorption decay of lower channeling modes to produce angular 
collimation of X-ray beams.
Estimates were given for optimum capillary length at different
roughness amplitudes for angular sensitivity of X-ray transmission
and chenneling effects that can be useful for designing of detector
systems.
}

\newpage

Capture of X-ray beam into a narrow dielectric capillary reviels many 
features of mode behavior typical for cannelling of relativistic 
positrons in crystals \cite{ 2, 3}. 
X-ray scattring at rough surfaces is usually investigated within 
the well known Andronov-Leontovich approach \cite{27}.
A model with transitional layer was proposed in \cite{28}.
Within both the models scattering and absorption disappear
at small grazing angle limit that appearently results from assumption
of plane wave incidence that is not true for submicron capillaries.
The evolution of the 
chenneled X-ray beam can
be calculated either by direct integration of the wave equation
\cite{8} or on the basis of the mode approach. 
In the latter
case the account for strong incoherent scattering is serious problem.

It was shown that at small correlation lengths of roughness
absorption can be prevailing effect. Estimates are given for optimum 
capillary length at different roughness amplitudes  for angular 
sensitivity of X-ray transmission and channeling effects that can 
be useful in designing of detector systems. 


A new approach was developed for 
description of propagation of X-ray beam in thin dielectric capillary
with rough absorbing walls on the basis of slowly varying scalar 
amplitudes of electrical field vector $A(x,y,z)$.
In this case large angle scattering is neglected so
$$
\partial^2 <A(x,y,z)>/ \partial z^2 \ll k \cdot \partial <A(x,y,z)>/ \partial z
$$
and because  the beam is narrow
$$
\partial^2 <A(x,y,z)>/ \partial z^2 \ll 
\partial^2 <A(x,y,z)>/ \partial x^2 , 
$$
where $z$ and $x$ are coordinates along and across the channel.
The assumption results in "parabolic equation" of quazioptics.
 
In the approach the statistical method of Tatarsky (see \cite{4})
was generalised to include stratified media similar to the case of 
electron channeling in single crystals \cite{6}.
The dielectric permitance on the rough boundary with the random shape
$x=\xi (z)$ was presented as
$$
\varepsilon(x,z)=\varepsilon_1 +(\varepsilon_0 - \varepsilon_1 )H(x-\xi(z))  
$$
where $\varepsilon_1$ and $\varepsilon_0$ is the dielectric permittance 
of the substance and of the air respectively, $H(x)$ is a step function.
So the dielectric permittance of the transitional layer is the function of 
the perpendicular to boundary coordinate $x$
can be written
$$
<\varepsilon(x)>=\varepsilon_1 +(\varepsilon_0 -\varepsilon_1) 
 \int^{x}_{-\infty }P(\xi )d \xi
$$
where $P(\xi)$ is 
the propability distribution of roughness heights
which is assumed to be normal.

The coherent part of the amplitude $A(x,y,z) $ can be  calculated 
from the statistically averaged equation (angular brackets 
correspond to averaging)
$$
\begin{array}{l}
2ik \partial <A(x,y,z)>/ \partial z - \Delta _{\perp }<A(x,y,z)> -\\
\qquad {}- k^{2} {\chi}(x,y)
 < A(x,y,z)> - 
ik^{2} W(x,y)< A(x,y,z) > = 0,
\end{array}
\eqno (1)
$$
\noindent
$$
A(x,y,z=0)=A_{0}(x,y),
$$
where
$$
{\chi}(x,y)={(<{\varepsilon} (x,y)> - {\varepsilon}_1)/{\varepsilon_1}}.
$$
Roughness is accounted for as
scattering potential $W(x,y)$ and can be compared with real absorption 
$Im({\chi}(x,y))$ at
various parameters of the capillary. The value of incoherent absorption 
term at the boundary can be written as 
$$
W (x) = - {k\over 4}{(\varepsilon_0 -\varepsilon_1)^2 
\over{\pi {(\varepsilon_0)}^2}}
\int^{+\infty }_{-\infty }dz^\prime
\int^{x/\sigma }_{-\infty } exp(-{\xi}^2) d{\xi}
$$
$$
\int^{{x/\sigma -R(z^\prime)\xi}\over{(1-R^2(z^\prime))^{1/2}} }_{x/\sigma } 
 exp(-{\eta}^2) d\eta 
\eqno (2)
$$
where $R(z)$ is the autocorrelation coefficient,
$\sigma$ is dispersion of ${\xi}(z)$ distribution.
It can be shown that the value $W(x)$ in the middle of the transitional 
layer $(x=0)$ does not depend on $\sigma$ and is nearly proportional
to the correlation length $\l_{corr}$ of roughness.
\par

\noindent The inner double integrals in equation (2) can be 
simplified in the limit
$R \ll 1$
$$
{1\over{2\pi}}R(z^\prime) exp(-x^2/{\sigma^2})
$$
so approximation for $W(x)$ can be written as
$$
W(x)\approx -{k \over 4} 
{
{(\varepsilon_0-\varepsilon_1)}^2 \over{\pi {(\varepsilon_0)}^2}
}
\int_{-\infty}^{\infty}dz^\prime  
\int_{-\infty}^{0} exp(-\xi^2)d\xi
\int_{0}^
{
{-R(z^\prime)\xi}\over{{(1-R^2(z^\prime))}^{1/2}}
}
exp(-{\eta}^2)d\eta
$$
$$
 {\cdot} exp(-{{x^2}\over{\sigma^2}}) 
\eqno (3)
$$
with clear dependence on vertical coordinate $x$.

Scattering potential $W(x)$ and normalized dielectric permittance 
$\chi(x)$ across a flat glass channel were calculated for X-ray 
energy $E=10 keV$,
$\sigma=100 \AA$ and correlation length of roughness 
$\l_{corr}=2\mu m$.

\noindent Normalized dielectric permittance $Re(\chi(x,y))$ and 
scattering potential $W(x)$ curves are shown on Fig. 1
in arbitrary units.

\vskip 20pt
\begin{center}
\begin{minipage}{11cm}

\setlength{\unitlength}{0.240900pt}
\ifx\plotpoint\undefined\newsavebox{\plotpoint}\fi
\sbox{\plotpoint}{\rule[-0.200pt]{0.400pt}{0.400pt}}%
\special{em:linewidth 0.4pt}%
\begin{picture}(1500,900)(0,0)
\font\gnuplot=cmr10 at 10pt
\gnuplot
\put(220,113){\special{em:moveto}}
\put(1436,113){\special{em:lineto}}
\put(828,113){\special{em:moveto}}
\put(828,832){\special{em:lineto}}
\put(220,113){\special{em:moveto}}
\put(240,113){\special{em:lineto}}
\put(1436,113){\special{em:moveto}}
\put(1416,113){\special{em:lineto}}
\put(198,113){\makebox(0,0)[r]{0}}
\put(220,185){\special{em:moveto}}
\put(240,185){\special{em:lineto}}
\put(1436,185){\special{em:moveto}}
\put(1416,185){\special{em:lineto}}
\put(198,185){\makebox(0,0)[r]{0.1}}
\put(220,257){\special{em:moveto}}
\put(240,257){\special{em:lineto}}
\put(1436,257){\special{em:moveto}}
\put(1416,257){\special{em:lineto}}
\put(198,257){\makebox(0,0)[r]{0.2}}
\put(220,329){\special{em:moveto}}
\put(240,329){\special{em:lineto}}
\put(1436,329){\special{em:moveto}}
\put(1416,329){\special{em:lineto}}
\put(198,329){\makebox(0,0)[r]{0.3}}
\put(220,401){\special{em:moveto}}
\put(240,401){\special{em:lineto}}
\put(1436,401){\special{em:moveto}}
\put(1416,401){\special{em:lineto}}
\put(198,401){\makebox(0,0)[r]{0.4}}
\put(220,473){\special{em:moveto}}
\put(240,473){\special{em:lineto}}
\put(1436,473){\special{em:moveto}}
\put(1416,473){\special{em:lineto}}
\put(198,473){\makebox(0,0)[r]{0.5}}
\put(220,544){\special{em:moveto}}
\put(240,544){\special{em:lineto}}
\put(1436,544){\special{em:moveto}}
\put(1416,544){\special{em:lineto}}
\put(198,544){\makebox(0,0)[r]{0.6}}
\put(220,616){\special{em:moveto}}
\put(240,616){\special{em:lineto}}
\put(1436,616){\special{em:moveto}}
\put(1416,616){\special{em:lineto}}
\put(198,616){\makebox(0,0)[r]{0.7}}
\put(220,688){\special{em:moveto}}
\put(240,688){\special{em:lineto}}
\put(1436,688){\special{em:moveto}}
\put(1416,688){\special{em:lineto}}
\put(198,688){\makebox(0,0)[r]{0.8}}
\put(220,760){\special{em:moveto}}
\put(240,760){\special{em:lineto}}
\put(1436,760){\special{em:moveto}}
\put(1416,760){\special{em:lineto}}
\put(198,760){\makebox(0,0)[r]{0.9}}
\put(220,832){\special{em:moveto}}
\put(240,832){\special{em:lineto}}
\put(1436,832){\special{em:moveto}}
\put(1416,832){\special{em:lineto}}
\put(198,832){\makebox(0,0)[r]{1}}
\put(220,113){\special{em:moveto}}
\put(220,133){\special{em:lineto}}
\put(220,832){\special{em:moveto}}
\put(220,812){\special{em:lineto}}
\put(220,68){\makebox(0,0){-400}}
\put(372,113){\special{em:moveto}}
\put(372,133){\special{em:lineto}}
\put(372,832){\special{em:moveto}}
\put(372,812){\special{em:lineto}}
\put(372,68){\makebox(0,0){-300}}
\put(524,113){\special{em:moveto}}
\put(524,133){\special{em:lineto}}
\put(524,832){\special{em:moveto}}
\put(524,812){\special{em:lineto}}
\put(524,68){\makebox(0,0){-200}}
\put(676,113){\special{em:moveto}}
\put(676,133){\special{em:lineto}}
\put(676,832){\special{em:moveto}}
\put(676,812){\special{em:lineto}}
\put(676,68){\makebox(0,0){-100}}
\put(828,113){\special{em:moveto}}
\put(828,133){\special{em:lineto}}
\put(828,832){\special{em:moveto}}
\put(828,812){\special{em:lineto}}
\put(828,68){\makebox(0,0){0}}
\put(980,113){\special{em:moveto}}
\put(980,133){\special{em:lineto}}
\put(980,832){\special{em:moveto}}
\put(980,812){\special{em:lineto}}
\put(980,68){\makebox(0,0){100}}
\put(1132,113){\special{em:moveto}}
\put(1132,133){\special{em:lineto}}
\put(1132,832){\special{em:moveto}}
\put(1132,812){\special{em:lineto}}
\put(1132,68){\makebox(0,0){200}}
\put(1284,113){\special{em:moveto}}
\put(1284,133){\special{em:lineto}}
\put(1284,832){\special{em:moveto}}
\put(1284,812){\special{em:lineto}}
\put(1284,68){\makebox(0,0){300}}
\put(1436,113){\special{em:moveto}}
\put(1436,133){\special{em:lineto}}
\put(1436,832){\special{em:moveto}}
\put(1436,812){\special{em:lineto}}
\put(1436,68){\makebox(0,0){400}}
\put(220,113){\special{em:moveto}}
\put(1436,113){\special{em:lineto}}
\put(1436,832){\special{em:lineto}}
\put(220,832){\special{em:lineto}}
\put(220,113){\special{em:lineto}}
\put(45,472){\makebox(0,0){$a. u.$}}
\put(828,23){\makebox(0,0){$x, \AA$}}
\put(828,877){\makebox(0,0){Fig.1}}
\sbox{\plotpoint}{\rule[-0.400pt]{0.800pt}{0.800pt}}%
\special{em:linewidth 0.8pt}%
\put(1306,767){\makebox(0,0)[r]{$\chi(x)$}}
\put(1328,767){\special{em:moveto}}
\put(1394,767){\special{em:lineto}}
\put(235,113){\special{em:moveto}}
\put(250,113){\special{em:lineto}}
\put(266,113){\special{em:lineto}}
\put(281,113){\special{em:lineto}}
\put(296,113){\special{em:lineto}}
\put(311,113){\special{em:lineto}}
\put(326,113){\special{em:lineto}}
\put(342,113){\special{em:lineto}}
\put(357,114){\special{em:lineto}}
\put(372,114){\special{em:lineto}}
\put(387,114){\special{em:lineto}}
\put(402,115){\special{em:lineto}}
\put(418,115){\special{em:lineto}}
\put(433,116){\special{em:lineto}}
\put(448,117){\special{em:lineto}}
\put(463,119){\special{em:lineto}}
\put(478,121){\special{em:lineto}}
\put(494,123){\special{em:lineto}}
\put(509,126){\special{em:lineto}}
\put(524,129){\special{em:lineto}}
\put(539,134){\special{em:lineto}}
\put(554,139){\special{em:lineto}}
\put(570,145){\special{em:lineto}}
\put(585,152){\special{em:lineto}}
\put(600,161){\special{em:lineto}}
\put(615,171){\special{em:lineto}}
\put(630,183){\special{em:lineto}}
\put(646,196){\special{em:lineto}}
\put(661,211){\special{em:lineto}}
\put(676,227){\special{em:lineto}}
\put(691,245){\special{em:lineto}}
\put(706,265){\special{em:lineto}}
\put(722,287){\special{em:lineto}}
\put(737,310){\special{em:lineto}}
\put(752,335){\special{em:lineto}}
\put(767,361){\special{em:lineto}}
\put(782,388){\special{em:lineto}}
\put(798,415){\special{em:lineto}}
\put(813,444){\special{em:lineto}}
\put(828,473){\special{em:lineto}}
\put(843,501){\special{em:lineto}}
\put(858,530){\special{em:lineto}}
\put(874,557){\special{em:lineto}}
\put(889,584){\special{em:lineto}}
\put(904,610){\special{em:lineto}}
\put(919,635){\special{em:lineto}}
\put(934,658){\special{em:lineto}}
\put(950,680){\special{em:lineto}}
\put(965,700){\special{em:lineto}}
\put(980,718){\special{em:lineto}}
\put(995,734){\special{em:lineto}}
\put(1010,749){\special{em:lineto}}
\put(1026,762){\special{em:lineto}}
\put(1041,774){\special{em:lineto}}
\put(1056,784){\special{em:lineto}}
\put(1071,793){\special{em:lineto}}
\put(1086,800){\special{em:lineto}}
\put(1102,806){\special{em:lineto}}
\put(1117,811){\special{em:lineto}}
\put(1132,816){\special{em:lineto}}
\put(1147,819){\special{em:lineto}}
\put(1162,822){\special{em:lineto}}
\put(1178,824){\special{em:lineto}}
\put(1193,826){\special{em:lineto}}
\put(1208,828){\special{em:lineto}}
\put(1223,829){\special{em:lineto}}
\put(1238,829){\special{em:lineto}}
\put(1254,830){\special{em:lineto}}
\put(1269,831){\special{em:lineto}}
\put(1284,831){\special{em:lineto}}
\put(1299,831){\special{em:lineto}}
\put(1314,831){\special{em:lineto}}
\put(1330,832){\special{em:lineto}}
\put(1345,832){\special{em:lineto}}
\put(1360,832){\special{em:lineto}}
\put(1375,832){\special{em:lineto}}
\put(1390,832){\special{em:lineto}}
\put(1406,832){\special{em:lineto}}
\put(1421,832){\special{em:lineto}}
\put(1436,832){\special{em:lineto}}
\sbox{\plotpoint}{\rule[-0.500pt]{1.000pt}{1.000pt}}%
\special{em:linewidth 1.0pt}%
\put(1306,722){\makebox(0,0)[r]{$W(x)$}}
\multiput(1328,722)(20.756,0.000){4}{\usebox{\plotpoint}}
\put(1394,722){\usebox{\plotpoint}}
\put(235,113){\usebox{\plotpoint}}
\put(235.00,113.00){\usebox{\plotpoint}}
\put(255.76,113.00){\usebox{\plotpoint}}
\put(276.51,113.00){\usebox{\plotpoint}}
\multiput(281,113)(20.756,0.000){0}{\usebox{\plotpoint}}
\put(297.27,113.00){\usebox{\plotpoint}}
\put(318.02,113.00){\usebox{\plotpoint}}
\put(338.78,113.00){\usebox{\plotpoint}}
\multiput(342,113)(20.756,0.000){0}{\usebox{\plotpoint}}
\put(359.53,113.00){\usebox{\plotpoint}}
\put(380.29,113.00){\usebox{\plotpoint}}
\put(401.04,113.00){\usebox{\plotpoint}}
\multiput(402,113)(20.756,0.000){0}{\usebox{\plotpoint}}
\put(421.79,113.25){\usebox{\plotpoint}}
\put(442.52,114.00){\usebox{\plotpoint}}
\multiput(448,114)(20.710,1.381){0}{\usebox{\plotpoint}}
\put(463.24,115.03){\usebox{\plotpoint}}
\put(483.82,117.73){\usebox{\plotpoint}}
\put(504.29,121.06){\usebox{\plotpoint}}
\multiput(509,122)(20.055,5.348){0}{\usebox{\plotpoint}}
\put(524.40,126.16){\usebox{\plotpoint}}
\put(543.31,134.59){\usebox{\plotpoint}}
\put(560.64,145.98){\usebox{\plotpoint}}
\put(576.19,159.60){\usebox{\plotpoint}}
\put(589.72,175.30){\usebox{\plotpoint}}
\multiput(600,189)(10.679,17.798){2}{\usebox{\plotpoint}}
\put(621.78,228.47){\usebox{\plotpoint}}
\multiput(630,246)(8.238,19.051){2}{\usebox{\plotpoint}}
\multiput(646,283)(6.697,19.645){3}{\usebox{\plotpoint}}
\multiput(661,327)(5.857,19.912){2}{\usebox{\plotpoint}}
\multiput(676,378)(5.461,20.024){3}{\usebox{\plotpoint}}
\multiput(691,433)(5.114,20.116){3}{\usebox{\plotpoint}}
\multiput(706,492)(5.266,20.076){3}{\usebox{\plotpoint}}
\multiput(722,553)(4.881,20.173){3}{\usebox{\plotpoint}}
\multiput(737,615)(5.197,20.094){3}{\usebox{\plotpoint}}
\multiput(752,673)(5.652,19.971){3}{\usebox{\plotpoint}}
\multiput(767,726)(6.697,19.645){2}{\usebox{\plotpoint}}
\multiput(782,770)(8.838,18.780){2}{\usebox{\plotpoint}}
\put(809.37,819.92){\usebox{\plotpoint}}
\put(826.15,831.14){\usebox{\plotpoint}}
\multiput(828,832)(18.808,-8.777){0}{\usebox{\plotpoint}}
\multiput(843,825)(12.064,-16.889){2}{\usebox{\plotpoint}}
\put(865.61,787.84){\usebox{\plotpoint}}
\multiput(874,770)(6.697,-19.645){3}{\usebox{\plotpoint}}
\multiput(889,726)(5.652,-19.971){2}{\usebox{\plotpoint}}
\multiput(904,673)(5.197,-20.094){3}{\usebox{\plotpoint}}
\multiput(919,615)(4.881,-20.173){3}{\usebox{\plotpoint}}
\multiput(934,553)(5.266,-20.076){3}{\usebox{\plotpoint}}
\multiput(950,492)(5.114,-20.116){3}{\usebox{\plotpoint}}
\multiput(965,433)(5.461,-20.024){3}{\usebox{\plotpoint}}
\multiput(980,378)(5.857,-19.912){3}{\usebox{\plotpoint}}
\multiput(995,327)(6.697,-19.645){2}{\usebox{\plotpoint}}
\multiput(1010,283)(8.238,-19.051){2}{\usebox{\plotpoint}}
\put(1032.48,232.18){\usebox{\plotpoint}}
\multiput(1041,214)(10.679,-17.798){2}{\usebox{\plotpoint}}
\put(1063.82,178.57){\usebox{\plotpoint}}
\put(1077.01,162.59){\usebox{\plotpoint}}
\put(1092.09,148.43){\usebox{\plotpoint}}
\put(1109.18,136.69){\usebox{\plotpoint}}
\put(1127.80,127.68){\usebox{\plotpoint}}
\multiput(1132,126)(20.055,-5.348){0}{\usebox{\plotpoint}}
\put(1147.70,121.86){\usebox{\plotpoint}}
\put(1168.12,118.23){\usebox{\plotpoint}}
\put(1188.71,115.57){\usebox{\plotpoint}}
\multiput(1193,115)(20.710,-1.381){0}{\usebox{\plotpoint}}
\put(1209.39,114.00){\usebox{\plotpoint}}
\put(1230.13,113.52){\usebox{\plotpoint}}
\put(1250.87,113.00){\usebox{\plotpoint}}
\multiput(1254,113)(20.756,0.000){0}{\usebox{\plotpoint}}
\put(1271.62,113.00){\usebox{\plotpoint}}
\put(1292.38,113.00){\usebox{\plotpoint}}
\put(1313.13,113.00){\usebox{\plotpoint}}
\multiput(1314,113)(20.756,0.000){0}{\usebox{\plotpoint}}
\put(1333.89,113.00){\usebox{\plotpoint}}
\put(1354.64,113.00){\usebox{\plotpoint}}
\multiput(1360,113)(20.756,0.000){0}{\usebox{\plotpoint}}
\put(1375.40,113.00){\usebox{\plotpoint}}
\put(1396.16,113.00){\usebox{\plotpoint}}
\put(1416.91,113.00){\usebox{\plotpoint}}
\multiput(1421,113)(20.756,0.000){0}{\usebox{\plotpoint}}
\end{picture}

\vskip 15pt
\centerline{Fig. 1}
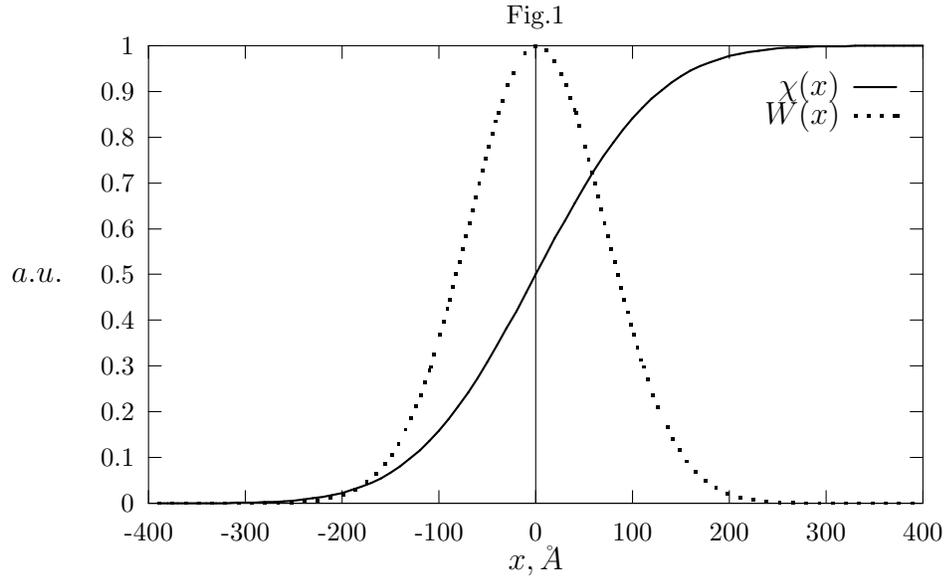
\noindent{\small Normalized dielectric permittance $Re(\chi(x,y))$ 
(solid line) and 
scattering potential $W(x)$ (dashed) curves}. 
\end{minipage}
\end{center}

Wave amplitude can be presented as a sum of modes
or eigenfunctions ${\varphi_j}(x,y)$
$$
A(x,y,z)=\sum_j C_j {\varphi_j}(x,y) exp(-ik_{jz}z -{\beta}_{jz}z)
\eqno (4)
$$
where eigenfunctions are solutions of equations
$$
\Delta_{\perp} {\varphi_j}(x,y)
=k[2k_{jz}-k Re(\chi(x,y))]{\varphi_j}(x,y).
\eqno (5)
$$
So attenuation coefficients can be found as overlap integrals
\par
\noindent
$$
\beta_l= -{k \over 2} \int{{\varphi_l}^\ast(x,y)
[Im(\chi(x,y))+W(x,y)]{\varphi_l}(x,y) }dxdy.
\eqno (6)
$$

It can be shown for lower channeled modes that incoherent scattering 
attenuation coefficient is proportional to $\sigma$ 

$$
\beta_{scatter} 
\sim 
k^2 {(\varepsilon_0-\varepsilon_1)}^2 \sigma
\int_{-\infty}^{\infty}dz^\prime  
\int_{-\infty}^{0} exp(-\xi^2/2)d\xi
\int_{0}^
{
{-R(z^\prime)\xi}\over{{(1-R^2(z^\prime))}^{1/2}}
}
exp(-{\eta/2}^2)d\eta.
$$

The results for attenuation coefficients [${\mu m}^{-1}$] for various 
wave modes are shown on Figure 2 separately for absorption and 
incoherent scattering within 0.5$\mu m$ quartz glass channel.

\vskip 20pt
\begin{center}
\begin{minipage}{11cm}

\setlength{\unitlength}{0.240900pt}
\ifx\plotpoint\undefined\newsavebox{\plotpoint}\fi
\sbox{\plotpoint}{\rule[-0.200pt]{0.400pt}{0.400pt}}%
\special{em:linewidth 0.4pt}%
\begin{picture}(1500,900)(0,0)
\font\gnuplot=cmr10 at 10pt
\gnuplot
\put(220,113){\special{em:moveto}}
\put(220,832){\special{em:lineto}}
\put(220,130){\special{em:moveto}}
\put(230,130){\special{em:lineto}}
\put(1436,130){\special{em:moveto}}
\put(1426,130){\special{em:lineto}}
\put(220,143){\special{em:moveto}}
\put(240,143){\special{em:lineto}}
\put(1436,143){\special{em:moveto}}
\put(1416,143){\special{em:lineto}}
\put(198,143){\makebox(0,0)[r]{1e-07}}
\put(220,183){\special{em:moveto}}
\put(230,183){\special{em:lineto}}
\put(1436,183){\special{em:moveto}}
\put(1426,183){\special{em:lineto}}
\put(220,235){\special{em:moveto}}
\put(230,235){\special{em:lineto}}
\put(1436,235){\special{em:moveto}}
\put(1426,235){\special{em:lineto}}
\put(220,262){\special{em:moveto}}
\put(230,262){\special{em:lineto}}
\put(1436,262){\special{em:moveto}}
\put(1426,262){\special{em:lineto}}
\put(220,275){\special{em:moveto}}
\put(240,275){\special{em:lineto}}
\put(1436,275){\special{em:moveto}}
\put(1416,275){\special{em:lineto}}
\put(198,275){\makebox(0,0)[r]{1e-06}}
\put(220,315){\special{em:moveto}}
\put(230,315){\special{em:lineto}}
\put(1436,315){\special{em:moveto}}
\put(1426,315){\special{em:lineto}}
\put(220,368){\special{em:moveto}}
\put(230,368){\special{em:lineto}}
\put(1436,368){\special{em:moveto}}
\put(1426,368){\special{em:lineto}}
\put(220,395){\special{em:moveto}}
\put(230,395){\special{em:lineto}}
\put(1436,395){\special{em:moveto}}
\put(1426,395){\special{em:lineto}}
\put(220,408){\special{em:moveto}}
\put(240,408){\special{em:lineto}}
\put(1436,408){\special{em:moveto}}
\put(1416,408){\special{em:lineto}}
\put(198,408){\makebox(0,0)[r]{1e-05}}
\put(220,448){\special{em:moveto}}
\put(230,448){\special{em:lineto}}
\put(1436,448){\special{em:moveto}}
\put(1426,448){\special{em:lineto}}
\put(220,500){\special{em:moveto}}
\put(230,500){\special{em:lineto}}
\put(1436,500){\special{em:moveto}}
\put(1426,500){\special{em:lineto}}
\put(220,527){\special{em:moveto}}
\put(230,527){\special{em:lineto}}
\put(1436,527){\special{em:moveto}}
\put(1426,527){\special{em:lineto}}
\put(220,540){\special{em:moveto}}
\put(240,540){\special{em:lineto}}
\put(1436,540){\special{em:moveto}}
\put(1416,540){\special{em:lineto}}
\put(198,540){\makebox(0,0)[r]{0.0001}}
\put(220,580){\special{em:moveto}}
\put(230,580){\special{em:lineto}}
\put(1436,580){\special{em:moveto}}
\put(1426,580){\special{em:lineto}}
\put(220,633){\special{em:moveto}}
\put(230,633){\special{em:lineto}}
\put(1436,633){\special{em:moveto}}
\put(1426,633){\special{em:lineto}}
\put(220,660){\special{em:moveto}}
\put(230,660){\special{em:lineto}}
\put(1436,660){\special{em:moveto}}
\put(1426,660){\special{em:lineto}}
\put(220,673){\special{em:moveto}}
\put(240,673){\special{em:lineto}}
\put(1436,673){\special{em:moveto}}
\put(1416,673){\special{em:lineto}}
\put(198,673){\makebox(0,0)[r]{0.001}}
\put(220,712){\special{em:moveto}}
\put(230,712){\special{em:lineto}}
\put(1436,712){\special{em:moveto}}
\put(1426,712){\special{em:lineto}}
\put(220,765){\special{em:moveto}}
\put(230,765){\special{em:lineto}}
\put(1436,765){\special{em:moveto}}
\put(1426,765){\special{em:lineto}}
\put(220,792){\special{em:moveto}}
\put(230,792){\special{em:lineto}}
\put(1436,792){\special{em:moveto}}
\put(1426,792){\special{em:lineto}}
\put(220,805){\special{em:moveto}}
\put(240,805){\special{em:lineto}}
\put(1436,805){\special{em:moveto}}
\put(1416,805){\special{em:lineto}}
\put(198,805){\makebox(0,0)[r]{0.01}}
\put(220,113){\special{em:moveto}}
\put(220,133){\special{em:lineto}}
\put(220,832){\special{em:moveto}}
\put(220,812){\special{em:lineto}}
\put(220,68){\makebox(0,0){0}}
\put(416,113){\special{em:moveto}}
\put(416,133){\special{em:lineto}}
\put(416,832){\special{em:moveto}}
\put(416,812){\special{em:lineto}}
\put(416,68){\makebox(0,0){5}}
\put(612,113){\special{em:moveto}}
\put(612,133){\special{em:lineto}}
\put(612,832){\special{em:moveto}}
\put(612,812){\special{em:lineto}}
\put(612,68){\makebox(0,0){10}}
\put(808,113){\special{em:moveto}}
\put(808,133){\special{em:lineto}}
\put(808,832){\special{em:moveto}}
\put(808,812){\special{em:lineto}}
\put(808,68){\makebox(0,0){15}}
\put(1005,113){\special{em:moveto}}
\put(1005,133){\special{em:lineto}}
\put(1005,832){\special{em:moveto}}
\put(1005,812){\special{em:lineto}}
\put(1005,68){\makebox(0,0){20}}
\put(1201,113){\special{em:moveto}}
\put(1201,133){\special{em:lineto}}
\put(1201,832){\special{em:moveto}}
\put(1201,812){\special{em:lineto}}
\put(1201,68){\makebox(0,0){25}}
\put(1397,113){\special{em:moveto}}
\put(1397,133){\special{em:lineto}}
\put(1397,832){\special{em:moveto}}
\put(1397,812){\special{em:lineto}}
\put(1397,68){\makebox(0,0){30}}
\put(220,113){\special{em:moveto}}
\put(1436,113){\special{em:lineto}}
\put(1436,832){\special{em:lineto}}
\put(220,832){\special{em:lineto}}
\put(220,113){\special{em:lineto}}
\put(45,472){\makebox(0,0){${\mu m}^{-1}$}}
\put(828,23){\makebox(0,0){$N$}}
\put(828,877){\makebox(0,0){Fig.2}}
\put(1306,767){\makebox(0,0)[r]{$\beta_{absorp}$}}
\put(1328,767){\special{em:moveto}}
\put(1394,767){\special{em:lineto}}
\put(220,113){\special{em:moveto}}
\put(259,193){\special{em:lineto}}
\put(298,240){\special{em:lineto}}
\put(338,273){\special{em:lineto}}
\put(377,299){\special{em:lineto}}
\put(416,320){\special{em:lineto}}
\put(455,338){\special{em:lineto}}
\put(495,354){\special{em:lineto}}
\put(534,368){\special{em:lineto}}
\put(573,381){\special{em:lineto}}
\put(612,392){\special{em:lineto}}
\put(651,403){\special{em:lineto}}
\put(691,413){\special{em:lineto}}
\put(730,423){\special{em:lineto}}
\put(769,431){\special{em:lineto}}
\put(808,440){\special{em:lineto}}
\put(848,448){\special{em:lineto}}
\put(887,456){\special{em:lineto}}
\put(926,464){\special{em:lineto}}
\put(965,472){\special{em:lineto}}
\put(1005,479){\special{em:lineto}}
\put(1044,487){\special{em:lineto}}
\put(1083,494){\special{em:lineto}}
\put(1122,502){\special{em:lineto}}
\put(1161,510){\special{em:lineto}}
\put(1201,518){\special{em:lineto}}
\put(1240,526){\special{em:lineto}}
\put(1279,536){\special{em:lineto}}
\put(1318,547){\special{em:lineto}}
\put(1358,560){\special{em:lineto}}
\put(1397,578){\special{em:lineto}}
\put(1436,611){\special{em:lineto}}
\sbox{\plotpoint}{\rule[-0.500pt]{1.000pt}{1.000pt}}%
\special{em:linewidth 1.0pt}%
\put(1306,722){\makebox(0,0)[r]{$\beta_{scatter}$}}
\multiput(1328,722)(20.756,0.000){4}{\usebox{\plotpoint}}
\put(1394,722){\usebox{\plotpoint}}
\put(220,310){\usebox{\plotpoint}}
\multiput(220,310)(9.188,18.611){5}{\usebox{\plotpoint}}
\multiput(259,389)(13.254,15.973){3}{\usebox{\plotpoint}}
\multiput(298,436)(16.010,13.208){2}{\usebox{\plotpoint}}
\multiput(338,469)(17.270,11.513){2}{\usebox{\plotpoint}}
\multiput(377,495)(18.275,9.840){3}{\usebox{\plotpoint}}
\multiput(416,516)(18.845,8.698){2}{\usebox{\plotpoint}}
\multiput(455,534)(19.434,7.288){2}{\usebox{\plotpoint}}
\multiput(495,549)(19.535,7.013){2}{\usebox{\plotpoint}}
\multiput(534,563)(19.838,6.104){2}{\usebox{\plotpoint}}
\multiput(573,575)(19.976,5.634){2}{\usebox{\plotpoint}}
\multiput(612,586)(20.105,5.155){2}{\usebox{\plotpoint}}
\multiput(651,596)(20.249,4.556){2}{\usebox{\plotpoint}}
\put(710.61,609.02){\usebox{\plotpoint}}
\multiput(730,613)(20.332,4.171){2}{\usebox{\plotpoint}}
\multiput(769,621)(20.332,4.171){2}{\usebox{\plotpoint}}
\multiput(808,629)(20.445,3.578){2}{\usebox{\plotpoint}}
\multiput(848,636)(20.514,3.156){2}{\usebox{\plotpoint}}
\multiput(887,642)(20.429,3.667){2}{\usebox{\plotpoint}}
\multiput(926,649)(20.587,2.639){2}{\usebox{\plotpoint}}
\multiput(965,654)(20.526,3.079){2}{\usebox{\plotpoint}}
\multiput(1005,660)(20.587,2.639){2}{\usebox{\plotpoint}}
\multiput(1044,665)(20.514,3.156){2}{\usebox{\plotpoint}}
\multiput(1083,671)(20.647,2.118){2}{\usebox{\plotpoint}}
\put(1140.85,677.42){\usebox{\plotpoint}}
\multiput(1161,680)(20.652,2.065){2}{\usebox{\plotpoint}}
\multiput(1201,684)(20.587,2.639){2}{\usebox{\plotpoint}}
\multiput(1240,689)(20.647,2.118){2}{\usebox{\plotpoint}}
\multiput(1279,693)(20.647,2.118){2}{\usebox{\plotpoint}}
\multiput(1318,697)(20.652,2.065){2}{\usebox{\plotpoint}}
\multiput(1358,701)(20.647,2.118){2}{\usebox{\plotpoint}}
\multiput(1397,705)(20.647,2.118){2}{\usebox{\plotpoint}}
\end{picture}

\vskip 15pt

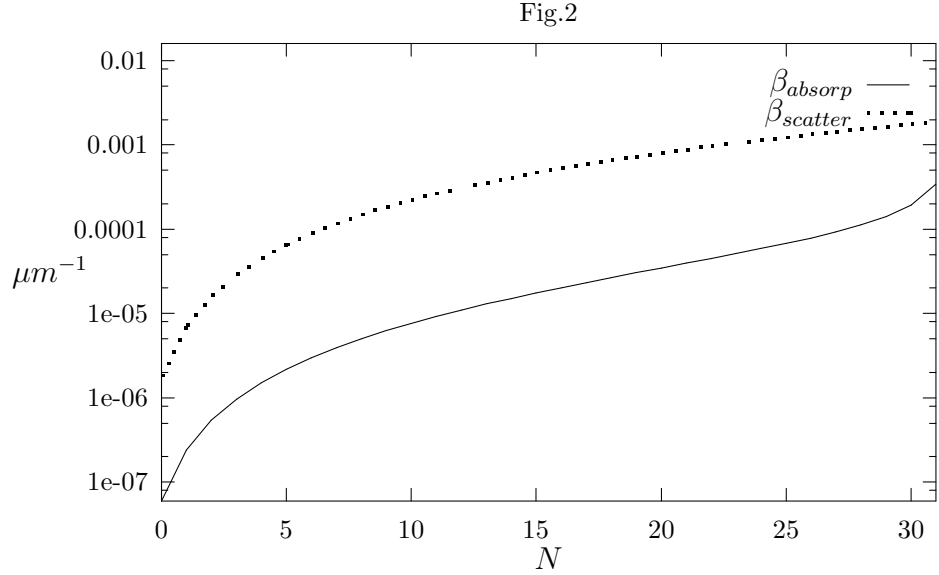
\centerline{Fig.~2}
\noindent {\small The dependence of attenuation coefficients for 
incoherent scattering 
${\beta}_{scatter}$ (dashed line) and absorption ${\beta}_{absorp}$ 
(solid line) on the mode number $N$.}
\end{minipage}
\end{center}
\vskip 20pt

As it is seen from the Figure 2 the effect of incoherent scattering
for given correlation length
is an order of magnitude higher than of real absorption. 
Coherent effects for the mode number "1" and higher will decay
after several $cm$ length of the capillary. And decreasing 
of correlation length
will result in nearly proportional decreasing of incoherent 
scattering.The effect can be measured
by the loss of angular dependence of outgoing beam on input beam
orientation for various quality surfaces.


\begin{thebibliography}{99}


\bibitem{2} {\sc M.J.~Pedersen, J.U.~Andersen, W.M.~Augustyniak}, 
 Radiation Effects, {\bf 12}, 47 (1972).

\bibitem{3} {\sc A.H.~S{\o}rensen, E.~Uggerh{\o}j}, Nuclear Science Applications,
  {\bf 3}, 147 (1989).

\bibitem{27} {\sc A.V.~Vinogradov, N.N.~Zorev, I.V.~Kozhevnikov, I.G.~Yakushkin},
ZhTF, {\bf 89}, 2124 (1985). 

\bibitem{28} {\sc A.V.~Andreev}, Uspehi Fiz. Nauk, {\bf 145},
113 (1985).


\bibitem{8} {\sc T.A.~Bobrova, L.I.~Ognev}, phys. stat. sol. (b), 
 {\bf 203/2}, R11 (1997).

\bibitem{4} {\sc V.~Hol\'y, K.T.Gabrielyan}, phys. stat. sol. (b), 
 {\bf 140}, 39 (1987).

\bibitem{6} {\sc L.I.~Ognev}, Radiation Effects  and  Defects  in  Solids, 
{\bf 25}, 81 (1993).



\end{thebibliography}
\end{document}